# A one femtojoule athermal silicon modulator

Erman Timurdogan, Cheryl M. Sorace-Agaskar, Jie Sun, Ehsan Shah Hosseini, Aleksandr Biberman and Michael R. Watts*

*Research Laboratory of Electronics, Massachusetts Institute of Technology, Cambridge, Massachusetts 02139, USA*

**Corresponding author email: mwatts@mit.edu*

**Silicon photonics has emerged as the leading candidate for implementing ultralow power wavelength division multiplexed communication networks in high-performance computers, yet current components (lasers, modulators, filters, and detectors) consume too much power for the femtojoule-class links that will ultimately be required. Here, we propose, demonstrate, and characterize the first modulator to achieve simultaneous high-speed (25-Gb/s), low voltage (0.5$V_{PP}$) and efficient 1-fJ/bit error-free operation while maintaining athermal operation. Both the low energy and athermal operation were enabled by a record free-carrier accumulation/depletion response obtained in a vertical *p-n* junction device that at 250-pm/V (30-GHz/V) is up to ten times larger than prior demonstrations. Over a 7.5°C temperature range, the massive electro-optic response was used to compensate for thermal drift without increasing energy consumption and over a 10°C temperature range, increasing energy consumption by only 2-fJ/bit. The results represent a new paradigm in modulator development, one where thermal compensation is achieved electro-optically.**

Intra- and inter-chip interconnects, traditionally executed using parallel electrical links, are struggling to satisfy stringent bandwidth, density, power consumption, and cost requirements of computing and communication industries [1−3]. Presently, these limitations are becoming apparent in rapidly scaling massively-parallel computing systems such as data centers, employed for cloud computing, and the high-performance "supercomputers" utilized for large-scale scientific computation. Moreover, this growth in bandwidth is driven by the exponentially growth in transistor count density predicted by Moore's Law. To maintain balance between communications and computation, communications bandwidth is also growing exponentially. Since the physics of communications does not naturally enable Moore's Law scaling, the interconnection network is increasingly becoming the performance bottleneck. While the current bottlenecks within the switches and interconnects of large-scale computing systems prove an immediate need for the inclusion of inter-chip optical components, on-chip (i.e. intra-chip) communication links, where the bandwidth and efficiency requirements are far more severe, provide a long-term vision. An interconnecting wire that communicates one bit over an average distance on-chip of 1mm (~10% of the die size), consumes 100fJ/bit [4]. At the 8nm node, over one petabit-per-second of data transfer will be required on large microprocessor die, leading to 100W of communications power per chip, nearly ten times as much power as is acceptable in practice. Therefore, on-chip photonic interconnects must be driven into the femtojoule-per-bit regime with each component within the link consuming just a fraction of that total [2−4].

Optical communications based on spatially division multiplexed vertical cavity surface emitting laser (VCSEL) arrays already exist in high-performance computing systems [3]. VCSEL based links, which utilize multimode fibers, meet requirements in terms of cost and power consumption at low data rates (~100Gb/s). However, they are not readily scalable to high-bandwidth (≥1Tb/s) operation with high efficiency due to their single-channel-per-fiber nature and multimode operation. Wavelength-division-multiplexed (WDM) systems offer high-bandwidth operation per fiber, and can be efficiently implemented by cascading multiple laser lines, encoded by on-chip silicon photonic modulators. Silicon photonics [5], inherently CMOS compatible, offers a high-index-contrast platform, leading to single-mode high-confinement waveguides and compact high-speed and low-energy consumption modulation [6]. Single-mode operation within this platform minimizes capacitance, enabling higher receiver sensitivity and lower overall power consumption.

In the silicon platform, high performance electro-absorptive germanium-on-silicon and electro-refractive silicon modulators have been demonstrated [7–20]. Germanium-on-silicon modulators utilize the strong Franz-Keldysh and Quantum-Confined-Stark effects, and have enabled low-power non-resonant modulators [19, 20]. However, electro-absorption devices also act as detectors and the photocurrent produced in the modulator with the required applied reverse bias leads to an additional significant term in the power consumption of the modulator [22]. Further, in order to achieve a large extinction ratio, germanium-on-silicon electro-absorption modulators to date have all required voltage levels that are not compatible with CMOS drive levels. Finally, since electro-absorption devices are based on a band edge effect, the material composition inherently limits the operation to a modest wavelength range. The band-edge can of course be engineered to enable wideband operation but at the expense of additional mask layers, regrowth, and fabrication complexity. In contrast, silicon modulators utilize the relatively weak free-carrier plasma dispersion effect [21], which alters the refractive index of silicon due to the change in polarization incurred by the change in carrier concentration. The index change is translated into a phase or frequency shift. The resultant shift can be modulated and translated into an amplitude response using a Mach-Zehnder interferometer or a resonator. Early demonstrations of silicon modulators on chip utilized poorly-confined ridge waveguides based on large Mach-Zehnder interferometers (>500μm in length) and were inefficient due to large drive voltage and/or current and device capacitance [7, 8]. In order to enhance the electro-refractive effect, high-confinement compact silicon resonant structures were proposed. Resonant modulators confine light in compact high quality factor (Q) devices that effectively increase the optical path length and the interaction with the change in the depletion width, where the free carrier concentration is modulated. Compact devices also minimize the device capacitance, and, thus, enable high-speed and low power operation. Q. Xu *et al.* [9] demonstrated the first silicon resonant electro-optic modulator based on carrier injection, which was then succeeded by a variety of silicon resonant modulators [11–18]. Injection based modulators have achieved large frequency shifts, but at the expense of low-speed operation (~1.5Gb/s) due to the long free-carrier lifetime within silicon diodes. High-speed operation can be achieved with the aid of signal pre-emphasis only at the expense of increased power consumption and CMOS complexity [10]. On the other hand, resonant silicon modulators based on the depletion of electrons and holes have achieved low-energy, low-voltage and high-bandwidth operation [11–18]. Initial ridge-based microring modulators based on lateral junctions with interior and exterior contacts, have achieved operation at 10Gb/s, an energy-per-bit of ~50fJ/bit [11] and an electro-optic response of 20pm/V (~2.5GHz/V) [11]. Recently, ridge-based microring modulators based on lateral and interleaved junctions with interior and exterior contacts, have achieved operation at 25Gb/s, an energy-per-bit of ~7fJ/bit [12] and 471fJ/bit [13] and >66fJ/bit [14], an electro-optic response of 26pm/V (~3GHz/V) [12] and 34pm/V (~4GHz/V) [13] and 40pm/V (~5GHz/V) [14], respectively. Vertical-junction microdisk modulators exhibit a much larger overlap of the depletion region with the optical mode, and as a result have achieved an electro-optic responses of ~60pm/V (7.5GHz/V) [15,16] and 90pm/V (11GHz/V) [17]. This substantially larger response has enabled ultra low energy-per-bit (3fJ/bit) modulation up to 12.5Gb/s [16] and, more recently, ~13fJ/bit modulation up to 25Gb/s [17]. In addition, the vertical junction enables interior p+ and n+ contacts that preserve the hard outer wall of the microdisk. The hard outer wall enables for a high confinement compact (~3.5µm diameter) silicon step-index microdisk, and a large free spectral range (FSR). A similar vertical junction microdisk modulator was driven differentially to further minimize the energy-per-bit down to 0.9fJ/bit at a data rate of 10Gb/s [18]. However, such results have been achieved only in thermally stable environments that are not representative of a real application. Within a microprocessor, baseline temperature variations across the processor are typically within ±7°C [23]. However, through barrel-shifting techniques [24], the temperature compensation range of each ring is set by the channel spacing. Implementation of this technique adds an energy per bit overhead of ~$A\times\log_2^N$, where N is the total number of resonators and A is the 2-by-1 electrical multiplexer energy consumption for any given node [25]. For 45nm, 32nm, 22nm and 11nm CMOS nodes, the estimated energy consumption of this technique for a 64 resonant modulators with 53 active modulators and 11 redundant modulators will be 16.7fJ/bit 9.2fJ/bit, 4.0fJ/bit, and 1.2fJ/bit, respectively, regardless of the thermal variations [25]. The extra redundant rings are required to achieve 99.9% link yield given the magnitude of random process variations [26] and the strict 10°C tuning range limit, which will be further minimized with the introduction of CMOS nodes (8nm and beyond) [24]. Given a thermo-optic response in silicon of 10GHz/°C, for a 100GHz channel spacing, 10°C must be compensated. To date, over a 10°C temperature range, the lowest tuning energy that would be possible based on reported heater and modulator efficiencies are ~82fJ/bit ($7\mu W/GHz\times100GHz/10^{10}bits/s+12fJ/bit=82fJ/bit$) at 10Gb/s [27] and ~175fJ/bit ($42\mu W/GHz\times100GHz/(2.5\times10^{10}bits/s)+7fJ/bit=175fJ/bit$) at 25Gb/s [12], which were dominating the modulator energy budget rather than electronics to drive them.

In this work, we present the design and demonstration of an athermal resonant silicon modulator that is compatible with a complementary metal-oxide-semiconductor (CMOS) process, and achieves the lowest total energy-per-bit performance to date. This is made possible by combining interior circular contacts with a highly optimized vertical junction within a compact microdisk modulator. The circular contact enables a direct and short electrical path out to the vertical junction, minimizing the resistance and corresponding loss in depletion/accumulation response while maximizing the speed of the modulator. Additionally, the vertical junction profile achieves an optimum balance between the mode overlap, quality factor and capacitance of the modulator. As a result, we achieve a record 250pm/V (30GHz/V) electro-optic frequency response and, thus, simultaneous error-free, high-speed (25-Gb/s), low-voltage ($0.5V_{PP}$) and 1-fJ/bit operation. Importantly, in bandwidth limited modulators achieving 1fJ/bit operation at 25Gb/s requires an electro-optic shift that is 2.5-times larger than at 10Gb/s in order to preserve the insertion loss and extinction ratio. Given the square-root dependence of the depletion width with applied voltage, doing so requires 6.25-times the voltage, resulting in ~39-times more energy-per-bit ($E_{bit} = CV^2/4$) than is required at 10Gb/s. Yet, with such a large electro-optic response, we not only achieve 1fJ/bit operation, but also simultaneously improve on the insertion loss and extinction ratio. Further, we utilize the record electro-optic frequency response to tune the microdisk modulator by varying the DC bias point and counteract the thermal shift over a $10^0C$ (~100GHz) temperature range with almost no static energy consumption, enabling ~1-fJ/bit and 3-fJ/bit operation over 7.5°C and $10^{\circ}C$ temperature ranges, respectively, and presenting a new paradigm in resonant modulators, whereby thermal shifts are compensated by electro-optic control. Finally, while these demonstrations were all performed at 25Gb/s, the modulator demonstrates open eye-diagrams at data rates as high as 44Gb/s when driven with higher voltages.

In a resonant depletion mode modulator, the junction capacitance-per-unit volume dictates the change in charge-per-unit volume and corresponding frequency shift, $\Delta\omega$, for a given applied voltage, V, as shown in [28] and described by the following equation.

$$\Delta\omega \propto -\omega \frac{CV}{v_o}$$

where, ω, $v_o$, and C are the resonant frequency, volume of the resonator core, and junction capacitance, respectively. Our microdisk modulator structure leverages a vertical junction and interior circular contact to enable a hard outer resonator wall for a tightly confined and compact device, as depicted in Fig. 1a and shown in the scanning electron microscopy image in Fig. 1b. The vertical junction in our modulator serves to maximize the capacitance-per-unit volume and thereby ensures maximal frequency shift for a given applied voltage. However, while increasing the capacitance in the optically active region of the resonator is helpful, reducing the overall capacitance of the device is desired since the switching energy goes as $E_s=CV^2$ [16]. The use of an interior circular contact together with a hard outer resonator wall aids in doing just that by minimizing the overall resonator volume and concentrating the capacitance only over the optically active region of the junction. Importantly, the interior circular contact also minimizes the device resistance since it only exhibits radial and vertical resistive components, thereby enabling the full potential to be dropped across the optically active region of the junction, rather than across large azimuthal resistances, which limited the performance of prior demonstrations [15-16,18]. The vertical and radial resistances are illustrated and identified using an electrical connection diagram in Fig. 1c on the cross-section of the circularly contacted microdisk. Given the very short electrical paths to the vertical junction there is effectively no parasitic resistance to inhibit the electro-optic response, enabling a much larger electro-optic response than was previously possible. Together with the large capacitance-per-unit-volume and small overall capacitance, this modulator represents a significant advance over prior geometries.

The implants, including the fabrication process steps (see the Methods Section for details), were simulated using Synopsys Technology Computer Aided Design (TCAD) Sentaurus Process. The device response as a function of voltage was simulated using Sentaurus Device, and the carrier concentration was extracted and is shown in Fig. 1d. The implant energies for formation of the vertical junction implants were slightly adjusted to fit the experimental spectral response as a function of voltage (see the Methods Section for details). The depletion width can be approximated to be the distance between 50% of the peak *n* and *p* doping concentrations. Using this approximation, the depletion width is estimated to be 124nm, 132nm and 139nm for an applied voltage of 0.25V, 0V, and −0.25V, respectively (Fig. 1e). The initial depletion width was optimized to maximize the mode overlap, electrical bandwidth of the modulator and the optical internal quality factor (Q). Additionally, the highly doped $p+$ and $n+$ regions, which act as low resistance contacts, were placed 1μm away from the outer wall of the resonator in order to minimize the overlap with the cylindrical optical mode and ensure high-Q operation. The radial electric field component of the optical mode, simulated using a complex, full-vectorial and cylindrical finite difference mode-solver (FDM) [29], is overlapped with the microdisk structure in Fig. 1a to illustrate high modal overlap between the vertical depletion region and the optical mode. The 220-nm-thick silicon microdisk with its hard outer resonator wall supports a radiation-limited internal Q of over one-million for a diameter as small as 4-μm according to the cylindrical FDM simulation. Here, we chose a 4.8-μm diameter microdisk due to contact geometry and process design rule limitations, as opposed to radiation limits. At this diameter, the capacitance of the modulator (C=Q/V) was estimated to be ~17fF from AC Sentaurus device simulations. For an AC coupled $0.5V_{PP}$ drive voltage, this corresponds to an estimated switching energy ($E_s=CV^2$) of ~4.25fJ. In a non-return-to-zero (NRZ) pseudo-random bit sequence (PRBS) signal, the 0-0, 0-1, 1-0 and 1-1 transitions are equally probable (p(0-0)=p(0-1)=p(1-1)=p(1-0)=1/4), and the modulator only consumes energy from the source in the 0-1 transitions ($E_{0-1}=E_s$, $E_{0-0}=0$, $E_{1-1}=0$, $E_{1-0}=0$). Therefore, the estimated energy-per-bit ($E_{bit}=p(0-0)\times E_{0-0}+p(0-1)\times E_{0-1}+p(1-1)\times E_{1-1}+p(1-0)\times E_{1-0}=E_s/4=CV^2/4$) is 1.06fJ/bit. An electrical step function from −0.25V to 0.25V was applied and the resultant step response, obtained from Sentaurus device simulations, was fitted with an exponential rise function. The electrical 3dB bandwidth $\left(f_{3dB}^{el}=1/\left(2\pi RC\right)\right)$ is estimated to be ~35GHz.

A continuous-wave (CW) lightwave generated by a tunable laser source was coupled to the fundamental TE-mode of the on-chip silicon waveguide using a tapered single mode fiber. The through port spectral response as a function of applied DC bias was measured at ~26.5°C and the results are plotted in Fig. 1d alongside the numerical simulations (see the Methods Section for details). Frequency shifts of +8GHz and −7GHz, as a result of accumulation and depletion of the junction, were observed under an applied reverse bias of 0.25V and −0.25V, respectively, agreeing well with the theoretical predictions and revealing a record 250pm/V (30GHz/V) electro-optic response. The full width at half maximum bandwidth of the optical resonance $\left(f_{FWHM}\right)$ was measured to be ~28GHz, which inherently limits the optical 3dB bandwidth $\left(f_{3dB}^{opt}=\sqrt{\sqrt{2}-1}\times f_{FWHM}\right)$ [31] to ~18GHz. This limit excludes the electrical 3dB bandwidth. The electro-optic bandwidth $\left(f_{3dB}^{el-opt}\right)$, including electrical and optical limitations, can be determined using the equation below.

$$\left(\frac{1}{f_{3dB}^{el-opt}}\right)^2=\left(\frac{1}{f_{3dB}^{opt}}\right)^2+\left(\frac{1}{f_{3dB}^{el}}\right)^2$$

Therefore, the estimated electro-optic 3dB bandwidth is ~16GHz.

The electrical power and switching energy were measured using a time domain reflectometer (TDR) (see the Methods Section for details). The measured power consumption and switching energy are shown in Fig. 2c for an AC coupled 0.5$V_{PP}$ dropped across the modulator. The exact switching energy and energy-per-bit of the microdisk were experimentally extracted (see the Methods Section for details) to be 3.65fJ and 0.91fJ/bit, respectively, which is in good agreement with estimated values through simulations. The slight difference between theory and experimental results can be explained with slightly lower p+ and n+ implant activation than the simulations, which reduces the overall device capacitance. With both direct integration and advanced 3D integration techniques, the wire (and/or pad capacitance) can be quite low. Using, for example, direct integration, the total via/wire capacitance has been estimated to be 5fF [24] and using wafer bonding based 3D integrated through-oxide-vias (TOV) [32], the via capacitance has been measured to be ~2fF. Therefore with TOVs, the ground and signal vias will add a total of 4fF of capacitance, similar to the direct integration case. In either case, the parasitic capacitance will add ~0.3fJ/bit, a negligible contribution compared to the modulator energy consumption.

High-speed optical eye diagrams, shown in Fig. 3a, were measured using a high-speed electro-optic test setup (see the Methods Section for details) with a digital sampling oscilloscope at data rates of 10Gb/s, 15Gb/s, 20Gb/s, and 25Gb/s. The dynamic extinction ratio and insertion loss are shown below each eye diagram in Fig. 3a. The dynamic extinction ratio (ER) is defined as the ratio of the maximum data level, <1>, and the minimum data level, <0>, which was calculated from optical true "0" and true "1" or ER=10log(<1>/<0>). The insertion loss (IL) was calculated using the following equation,

IL=10log((<1>−<0>)/("1"−<0>)). Optical true "0" and true "1" corresponds to the amplified spontaneous emission (ASE) floor after the filter and off resonance transmission level at λ~1588nm, respectively. The modulator exhibits low insertion loss (~1.0dB) and high extinction ratio (6.2dB) at a data rate of 25Gb/s for 0.5$V_{PP}$ drive voltage. The increased insertion loss at high data rates can be explained by the electro-optic bandwidth limitation.

In order to further quantify the modulator performance, Bit-Error-Rate (BER) measurements, shown in Fig. 3b, were performed at data rates of 20Gb/s and 25Gb/s. For data rates up to 25Gb/s, the device achieves error-free operation (BER $<10^{-12}$) for a PRBS pattern length of $2^{31}-1$. We did not observe any pattern dependence from $2^{7}-1$ to $2^{31}-1$ PRBS due to the use of the depletion-mode operation and extremely low drive voltages and low power consumption of the device [33]. A commercial $LiNbO_3$ Mach-Zehnder modulator was similarly characterized for reference. The commercial modulator is rated to a 3dB bandwidth of 35GHz, and driven with an AC coupled 5.5$V_{PP}$ electrical signal. The power penalty was recorded as the received power difference at a BER of $10^{-12}$ between the silicon microdisk modulator and the commercial $LiNbO_3$ modulator (Fig. 3b). Data transmission with the silicon microdisk modulator was received with a power penalty of 3.0dB and 3.06dB at data rates of 20Gb/s and 25Gb/s, respectively. The increased power penalty at high data rates can be explained by the electro-optic bandwidth difference between the two modulators and phase chirp, introduced by the resonant modulator.

The electro-optic response of the modulator was characterized using an electrical signal source with a frequency range from DC to 50GHz. The electrical signal output was calibrated to achieve a consistent peak-to-peak drive voltage for the entire frequency span. The laser was then aligned to the most linear part of the resonance λ~1578.4nm and a 50m$V_{PP}$ sinusoidal signal applied to the microdisk modulator at frequencies spanning 500MHz to 45GHz. The through port power was measured with an external PIN-TIA receiver and the peak to peak AC level was recorded and plotted in Fig. 3c. The electro-optic 3dB bandwidth was experimentally measured to be ~21GHz, which agrees closely with our theoretical small-signal estimate. The theoretical estimate of 16GHz was calculated for on-resonance operation and the modulator electro-optic bandwidth was measured slightly off-resonance to preserve linearity and to achieve a faster response [34].

The baseline electro-optic performance of these modulators, achieving a 250pm/V (30GHz/V) electro-optic response, to enable 1fJ/bit operation at 25Gb/s represents a new standard for resonant modulators. More importantly, the electro-optic response is sufficiently large to compensate for thermal shifts up to 10°C. Here, we demonstrate this principle by varying the temperature of the chip from 20°C to 30°C (see the Methods Section for details) and compensating for that variation with electro-optic tuning achieved by varying the bias voltage from -2.2V to 0.4V across the modulator contacts. The through port spectral response as a function of applied DC bias was measured at ~26.5°C and the results are plotted in Fig. 4a. The modulator shifted ~100GHz for an applied voltage bias from 0.4V to -2.2V. The resonant modulator spectrum without and with thermal compensation based on an applied voltage bias over a temperature range of 20°C to 30°C is shown in Fig. 4b-c, respectively. The compensating frequency shift originates primarily from electro-optic tuning with a minor contribution from thermo-optic tuning due to high leakage current at large reverse bias, −2.2V, resulting from band-to-band tunneling between p+ and n+ doped regions. The electro-optic tuning is simply a switching action, which only consumes energy in charging transitions. Given a microprocessor thermal time constant, $\tau \gg$ 1ms [23], a worst case switching power of 69.6pW ($P_s=CV^2/\tau$=69.6fJ/1ms=69.6pW), as determined from TDR measurements (see the Methods Section for details), is required resulting in a negligible energy-per-bit of 0.28-aJ/bit ($E_{bit}$=69.6pW/25Gb/s=0.28aJ/bit). Up to 27.5°C, the leakage current is sufficiently low that it does not contribute to additional tuning energy. To achieve a full 10°C temperature range, a -2.2V bias is required, which resulted in a relatively large leakage current of 23μA, and consumed a static power of 50.6μW corresponding to an energy-per-bit of 2fJ/bit. The leakage current also contributed to a small thermo-optic resonance frequency shift of ~7.2GHz ($\Delta f \approx$IV/7μW/GHz=7.2GHz [27]). While tuning the resonance appropriately for the given temperature, the modulator was driven by a terminated 0.5$V_{PP}$ drive with a PRBS length of $2^{31}$-1 at a data rate of 25-Gb/s across the 10°C temperature range without seeing any degradation in eye-quality or error rate. The resulting high-speed optical eye diagrams, shown in Fig. 4e, were measured using a digital sampling scope at a data rate of 25Gb/s over a 10°C temperature range. The high extinction ratio and low insertion loss were preserved over the 10°C temperature range at 25Gb/s transmission. BER measurements, shown in Fig. 4d, were performed at a data rate of 25Gb/s and the modulator achieves error-free operation (BER <$10^{-12}$) for a PRBS pattern length of $2^{31}-1$ over the 10°C temperature range. The power penalty of the data transmission at a BER of $10^{-12}$ due to thermal variations was within ±0.14dB. The total energy-per-bit was determined to be <1fJ/bit, shown on Fig. 4f for each case, and 3fJ/bit for over 7.5°C and 10°C temperature ranges, respectively. All prior demonstrations of resonant silicon modulators required thermo-optic tuning due to their relatively small electro-optic frequency responses. However, implementing thermal tuning within modulators requires additional contacts, complicating and often degrading the modulator performance, and when tuned over a 10°C temperature range, the energy consumption of these devices increased substantially to ~82fJ/bit (7μW/GHz×100GHz/$10^{10}$bits/s+12fJ/bit=82fJ/bit) [27] and ~175fJ/bit (42μW/GHz×100GHz/(2.5×$10^{10}$bits/s)+7fJ/bit=175fJ/bit) [12] at data rates of 10Gb/s and 25Gb/s, respectively.

In order to demonstrate the limits of high-speed digital operation of the modulator, a voltage amplifier and a bias tee with a 3-dB bandwidth of 40GHz were attached to the electro-optic test setup (see the Methods Section for details) to deliver a 2.2$V_{PP}$ drive signal with −0.5V DC bias. High-speed optical eye diagrams, shown in Fig. 5, were measured with a digital sampling oscilloscope at data rates of up to 44Gb/s. The dynamic extinction ratio and insertion loss are shown below each eye diagram in Fig. 5. With low insertion loss (0.9dB) and high extinction ratio (8.0dB), the eye diagrams are still open at 44Gb/s. There are a couple of points to consider that enabled this result. The NRZ 44Gb/s data stream only requires ~22GHz (half of the bitrate) of bandwidth since the spectrum of the NRZ signal is proportional to $\text{sinc}^2(f)$, where $f$ is the frequency at the bit rate and the major spectral components are below half of the bit rate [35]. Also, the electro-optic 3dB bandwidth was measured to be 21GHz with a small-signal drive (50m$V_{PP}$). However, when the modulator is driven with higher voltages, -2.2V, the modulator bandwidth increases to 22.4GHz due to the reduction of average device capacitance, measured by TDR (see below and Fig. 4-f), from ~16fF to ~14fF. This stems from the fact that the device capacitance is reduced significantly when it is reverse-biased on account of the increased depletion width, also observed in [13]. Additionally, the modulator started to operate in the optically limiting regime at high drive voltages, also observed in [13]. In this regime, the optical resonance limited the data levels to the optical “0” and “1” level, which enabled relatively faster transitions. This demonstration shows that the modulator can provide on-demand bandwidth with a measured energy consumption of 17.4fJ/bit (see the Methods Section for details).

We have demonstrated an athermal silicon photonic modulator with the largest electro-optic response, lowest power, and lowest single-ended voltage demonstrated to date over a $10^{\circ}C$ temperature range. The record 250pm/V (30GHz/V) electro-optic response enables error-free 1-fJ/bit and 3-fJ/bit operation at a data rate of 25Gb/s and AC coupled drive of $0.5V_{PP}$ in a 4.8μm diameter microdisk modulator over $7.5^{\circ}C$ and $10^{\circ}C$ temperature ranges, respectively. Driven harder, the modulator demonstrates open eye-diagrams up to 44Gb/s. The electro-optic response of the circularly contacted vertical junction device achieved an electro-optic response up to ten times larger than that achieved with lateral junction based modulators, enabling both efficient modulation and the electro-optic control of the resonant frequency. The combined result improves upon the overall energy consumption of modulators by nearly two orders of magnitude. No other resonant modulator today is capable of achieving a similar result. Yet, these results represent only the beginning of electro-optic control of the resonant frequency. Future modulators and filters with more highly optimized junctions will clear the way for direct control of even greater temperature excursions with even less leakage current. While the modulator represents only one component of a silicon photonic link, as photonics becomes more closely integrated with CMOS, and as the bandwidth requirements on- and off-chip continue to rise, the power consumption of each component will be pressed to achieve new levels of efficiency in order to fit within the overall power envelope of the communication link. With their efficient, 1fJ/bit operation at the increasingly important 25Gb/s data rate, and with the effective elimination of heater power consumption, these results represent an important first step towards achieving femtojoule-per-bit class communication links, links that will prove critical for future on-chip applications.

## Methods

**Fabrication.** The microdisk modulator was fabricated in a 300-mm CMOS foundry with a 65-nm technology node, using silicon-on-insulator (SOI) wafers with a 0.225-μm top silicon layer and 2-μm buried oxide (BOX) layer for optical isolation. A full silicon etch was applied to form the silicon resonant microdisk and bus waveguides. An oxidization step followed to passivate the sidewalls, which also decreased the waveguide thickness to 0.22-μm. The vertical junction, centered ~110nm thickness, with arsenic (As) and boron difluoride ($BF_2$) implants with a designed dose of $4\times10^{13}cm^{-2}$, $2.8\times10^{13}cm^{-2}$, and energy of 380keV, 120keV, respectively. Through fitting the experimental spectral response of the modulator as a function of voltage, the arsenic (As) and boron difluoride ($BF_2$) implants were determined to be formed with an equivalent energy of 415keV, 90keV, respectively. The overlapping n+ and p+ doped regions in the center of the microdisk were formed by phosphorus and $BF_2$ implants with doses $5\times10^{15}cm^{-2}$, $3.5\times10^{15}cm^{-2}$, and energy of 180keV, 120keV, respectively. An oxide ($SiO_2$) layer with a total thickness of 3.6-μm was deposited to cover the modulator. Low resistance contacts to p+ and n+ doped regions were made by standard silicidation, copper-silicon circular and point contacts. Finally, the contacts were connected to on-chip ground-signal-ground (GSG) probing pads by two metal layers.

**Numerical Simulation.** The carrier distribution as a function of voltage, simulated by Sentaurus, was converted into an index distribution using the conversion equation in [16, 21]. The index distribution was fed into the FDM [29] to determine the complex propagation constant as a function of voltage. The complex propagation constant was then input to the transfer matrix of the resonant modulator [30] in order to theoretically predict the spectral response as a function of voltage.

**Time Domain Reflectometer measurement.** The absorbed power of the modulator was determined by subtracting two TDR measurements. First, the reflected voltage ($V_P^-$) from the transmission line, non-terminated GSG probe and pads ($Z_P$) was initially determined using the experimental setup, shown in Fig. 2a. The reflected voltage was then normalized to the emanating step voltage ($V_{in}$) to extract the frequency domain reflection coefficient, $S_{11}^p = \frac{V_P^-}{V_{in}} = \frac{Z_P - Z_0}{Z_P + Z_0}$, where, $Z_0$, 50Ω, is the characteristic impedance of the transmission line. The absorbed switching power in this path was determined by the following equation, $P_s^p = \left(1-\left(S_{11}^p\right)^2\right)V_{in}^2 / Z_0$. Second, the reflected voltage ($V_{PM}^-$) from this path and the modulator ($Z_M$), connected to the identical wiring, was determined using the experimental setup, shown in Fig. 2b. The normalized frequency domain reflection coefficient and absorbed switching power was determined by the following equations, $S_{11}^{pm} = \frac{V_{PM}^-}{V_{in}} = \frac{Z_{PM} - Z_0}{Z_{PM} + Z_0}$ and $P_s^{pm} = \left(1 - \left(S_{11}^{pm}\right)^2\right) V_{in}^2 / Z_0$, respectively. With close electrical and photonics integration these devices will not require long transmission lines, probing or large GSG pads. Therefore, the stand-alone modulator absorbed power is $P_s^m = \left(1 - \left(S_{11}^m\right)^2\right) V_{in}^2 / Z_0$, which can be calculated by subtracting the previously measured power consumptions, $P_s^m = P_s^{pm} - P_s^p = \left(\left(S_{11}^p\right)^2 - \left(S_{11}^{pm}\right)^2\right) V_{in}^2 / Z_0$. The switching power of the modulator ($P_s^m$) was integrated over time to obtain switching energy, $E_s = \int P_s^m \, dt$. The exact switching energy using the above equation was determined to be 3.65fJ (0.91fJ/bit). Alternatively, using the parallel equivalent impedance model, $\frac{1}{Z_{PM}} = \frac{1}{Z_P} + \frac{1}{Z_M}$, we can approximate the modulator reflection coefficient with the following equation, $\frac{S_{11}^{pm}}{S_{11}^p} = \frac{V_{PM}^-}{V_P^-} = \frac{Z_M - Z_0 - \frac{Z_0^2}{Z_P^2}\left(Z_P + Z_0\right)}{Z_M + Z_0 + \frac{Z_0^2}{Z_P^2}\left(Z_P + Z_0\right)} \cong \frac{Z_M - Z_0}{Z_M + Z_0} = S_{11}^m$, which was also performed in [16]. This approximation is accurate for $Z_P >> Z_0^2$, and overestimates the high frequency power consumption as $Z_P$ approaches to $Z_0^2$. The approximated switching energy using the above equations was determined to be 3.84fJ (0.96fJ/bit), which is in good agreement with the exact switching energy.

**High-speed electro-optic setup.** The electrical data, encoded using a pattern generator, was a non-return-to-zero (NRZ) on-off-keyed (OOK) signal, encoded with a $2^{31}-1$ pseudo-random-bit-sequence (PRBS). The output of the pattern generator was $0.5V_{PP}$ with +0.25V DC bias, and a DC block was attached to achieve an AC coupled $0.5V_{PP}$ drive. A 50Ω-terminated GSG probe was used to eliminate transmission line reflections and ensure that the voltage dropped across the modulator was $0.5V_{PP}$. This termination will not be required in a digital communications application, since a silicon modulator, integrated with electronics, will have on the order of 10-to-100μm of wiring and the transmission line effects, for $\lambda_{RF}$>2mm, will be negligible even at 44Gb/s operation. The laser line, coupled into the microdisk, was set to λ~1587.4nm and the through port output was then passed through an erbium doped fiber amplifier (EDFA) to overcome fiber-to-chip coupling losses. A tunable bandpass filter with a 1nm 3dB bandwidth was used to filter out the amplified spontaneous emission (ASE) of the EDFA. Additionally, a variable optical attenuator (VOA) was inserted before the high-speed *p-i-n* photodiode and transimpedance amplifier (PIN-TIA) receiver. The received data was then fed differentially to a bit-error-rate tester (BERT) for evaluation. No pre-emphasis or equalization of the signal was used in the measurements. The total fiber-to-fiber insertion loss was ~14.5dB, and the intensity before the EDFA was ~−9dBm.

**Temperature-controlled setup.**

In order to alter and record the chip temperature, a thermoelectric-cooler (TEC) was attached to the silicon substrate and a thermistor, placed adjacent to the silicon chip, was utilized to feedback temperature readout. Then, a commercial TEC proportional-integral-derivative (PID) controller was used to stabilize the chip temperature from $20^{\circ}C$ to $30^{\circ}C$ with $\pm 0.05^{\circ}C$ accuracy. The CW laser is set to $\lambda$~1587.6nm and fiber to chip coupling is optimized for each temperature set point due to the thermal expansion of TEC.

**Acknowledgements**

This work was supported by the Defense Advanced Research Projects Agency (DARPA) of the United States under the E-PHI project, grant no. HR0011-12-2-0007 and DARPA POEM award no. HR0011-11-C-0100. Cheryl Sorace-Agaskar acknowledges support from NSFGRP, No. 0645960.

**Author contributions**

E.T. and M.R.W. conceived the idea of the project. E.T. simulated and designed the microdisk modulator, laid out the mask, performed the experimental characterizations, analyzed the data and wrote the manuscript. C.M.S-A. did the doping and the vertical junction simulations and supported the data analysis. J. S. took the scanning electron microscopy image of the modulator. E.S.H. coordinated the layout. A.B. helped with the electro-optic test setup, and contributed to the writing of the manuscript. M.R.W. edited the manuscript and supervised the project.

**Competing financial interests**

The authors declare that they have no competing financial interests.

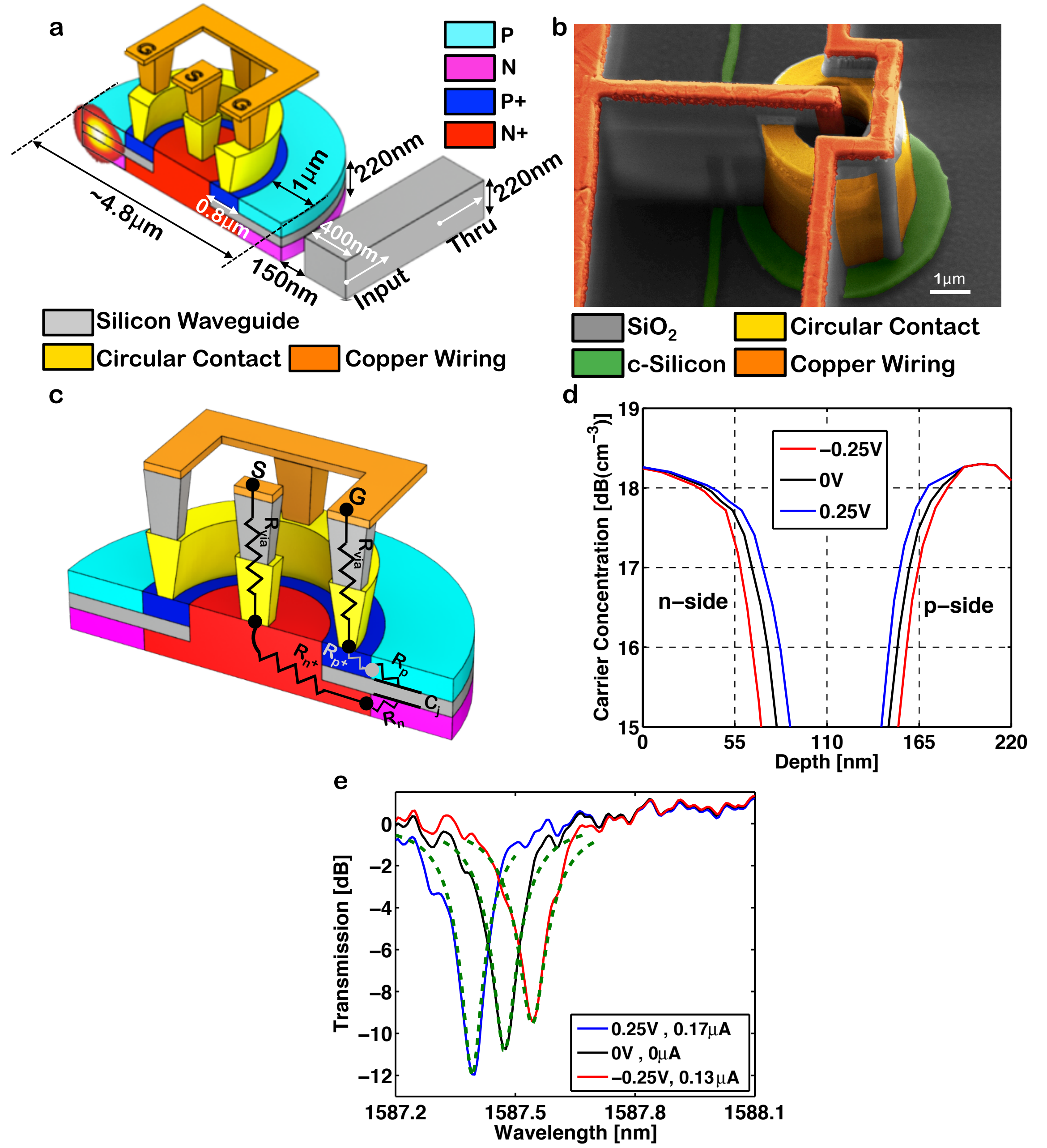

**Figure 1 | Silicon microdisk modulator and vertical junction design.** **a**, A 3D sketch of the electro-optic silicon microdisk modulator, showing the cross-section, size, doping profile, metal connections and the optical mode overlapped with the vertical *p-n* junction. **b,** A scanning electron microscopy image of the silicon microdisk modulator, revealed by dry etching the $SiO_2$ around the modulator to show the metal interconnect, circular contact, silicon bus waveguide and the microdisk. The signal pad, connected by short (~10-100μm) wires, is shown on the left side of the image. **c,** The electrical equivalent diagram of the circularly contacted microdisk, showing n+, p+, p, n doped and vertical via resistances ($R_{n+}$, $R_{p+}$,$R_p$ , $R_n$) as

well as the junction capacitance ($C_j$). Due to its cylindrical symmetry, the device has only vertical and radial resistance terms and no azimuthal resistance term, which significantly reduces the contact and device resistance, thereby enabling the full potential to be dropped across the optically active region of the junction and a resulting much larger electro-optic response. **d**, The simulated finite element model (FEM) vertical junction carrier distribution profile as a function of applied voltage. **e**, The measured and simulated (green dashed curves) transmission spectra of the resonator, overlaid on the experimental result, at ~26.5°C and with applied DC bias voltages of 0.25V, 0V and –0.25V, respectively.

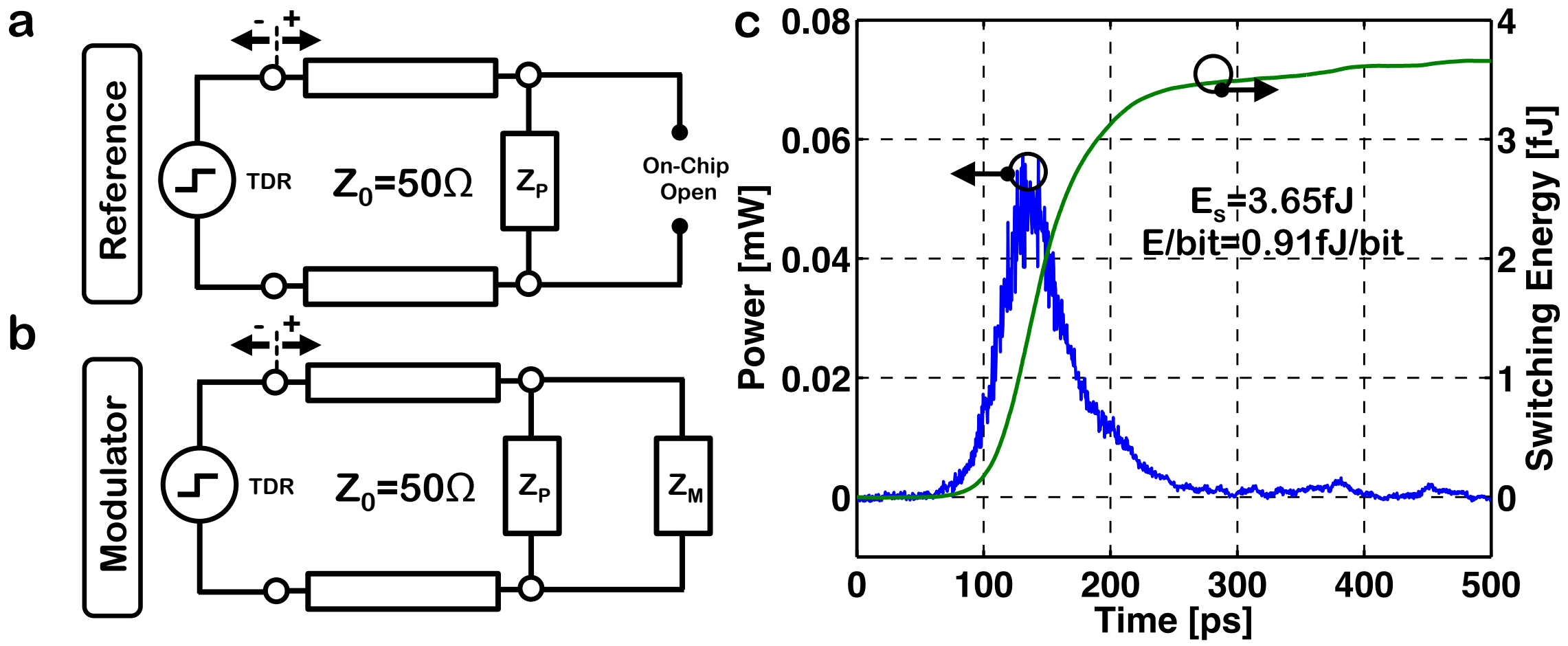

**Figure 2 | TDR measurement and switching energy.** **a**, The electrical setup for the time domain reflectometry (TDR) reference measurement, configured for measuring the reflected voltage ($V_P^-$) from the open ended on-chip pads ($Z_P$). **b,** The electrical setup for the time domain reflectometry (TDR) reference and modulator measurement, configured for measuring the reflected voltage ($V_{PM}^-$), from the modulator load ($Z_M$) and identical on-chip pads ($Z_P$). The TDR input and reflected TDR output directions are depicted with + and – signs. **c**, The time domain reflectometry (TDR) measurement of the electro-optic modulator power consumption and switching energy for a 0.5$V_{PP}$ AC-coupled drive.

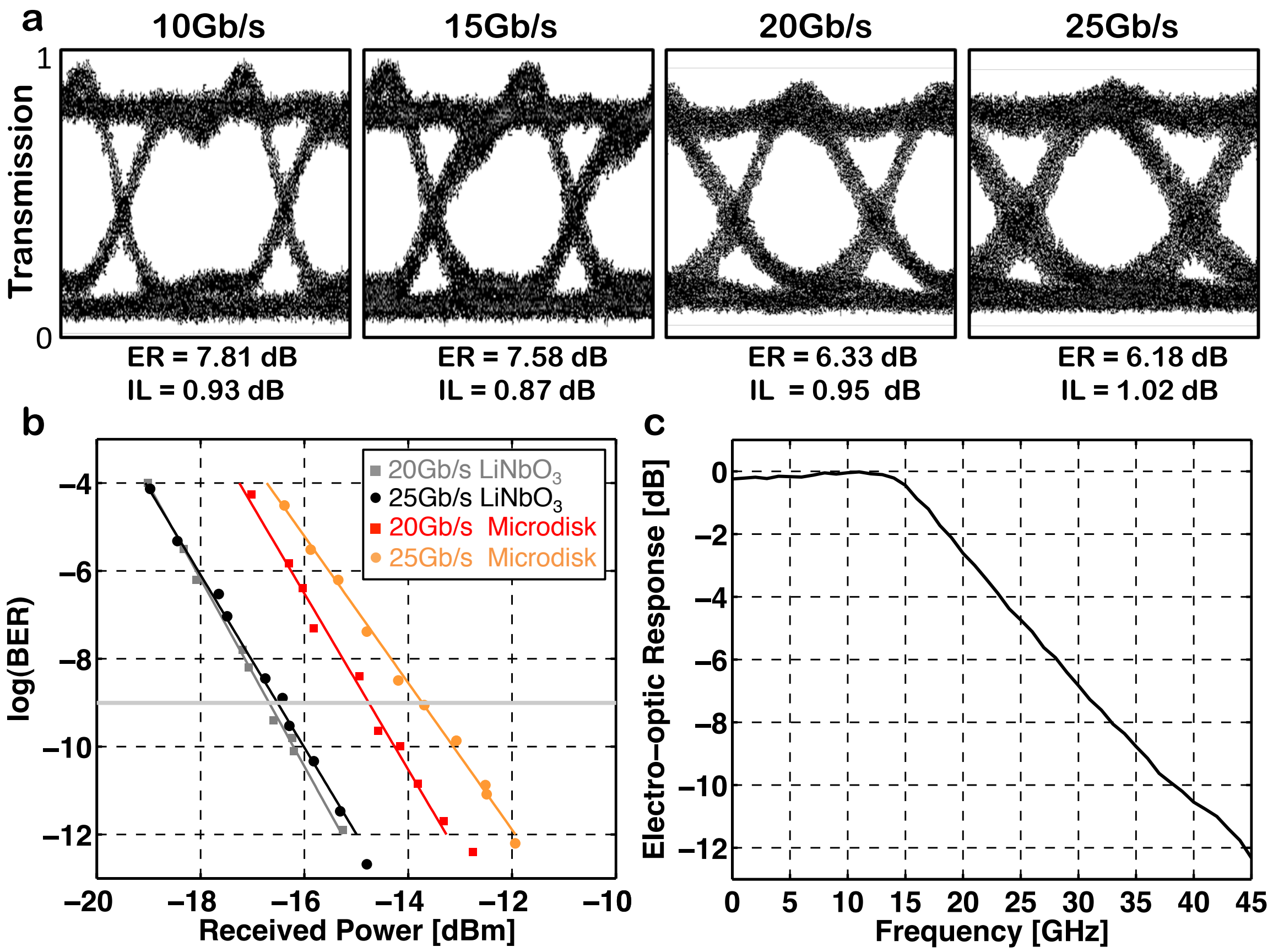

**Figure 3 | High speed modulator characterization. a**, The experimentally measured high-speed optical eye diagrams egressing from the silicon electro-optic modulator driven with a terminated probe and $0.5V_{PP}$ NRZ-OOK PRBS with a pattern length of $2^{31}$-1 at data rates of 10-, 15-, 20- and 25-Gb/s. The extinction ratio (ER) and insertion loss (IL) are denoted below the eye diagrams at each data rate. **b**, The bit-error-rate curves measured for the silicon microdisk and commercial $LiNbO_3$ modulators, for 20- and 25-Gb/s data rates. The bit-error-rate curves for the commercial $LiNbO_3$ modulator were used as a reference to obtain the power penalty of the silicon microdisk modulator. **c**, The electro-optic frequency response of the silicon modulator.

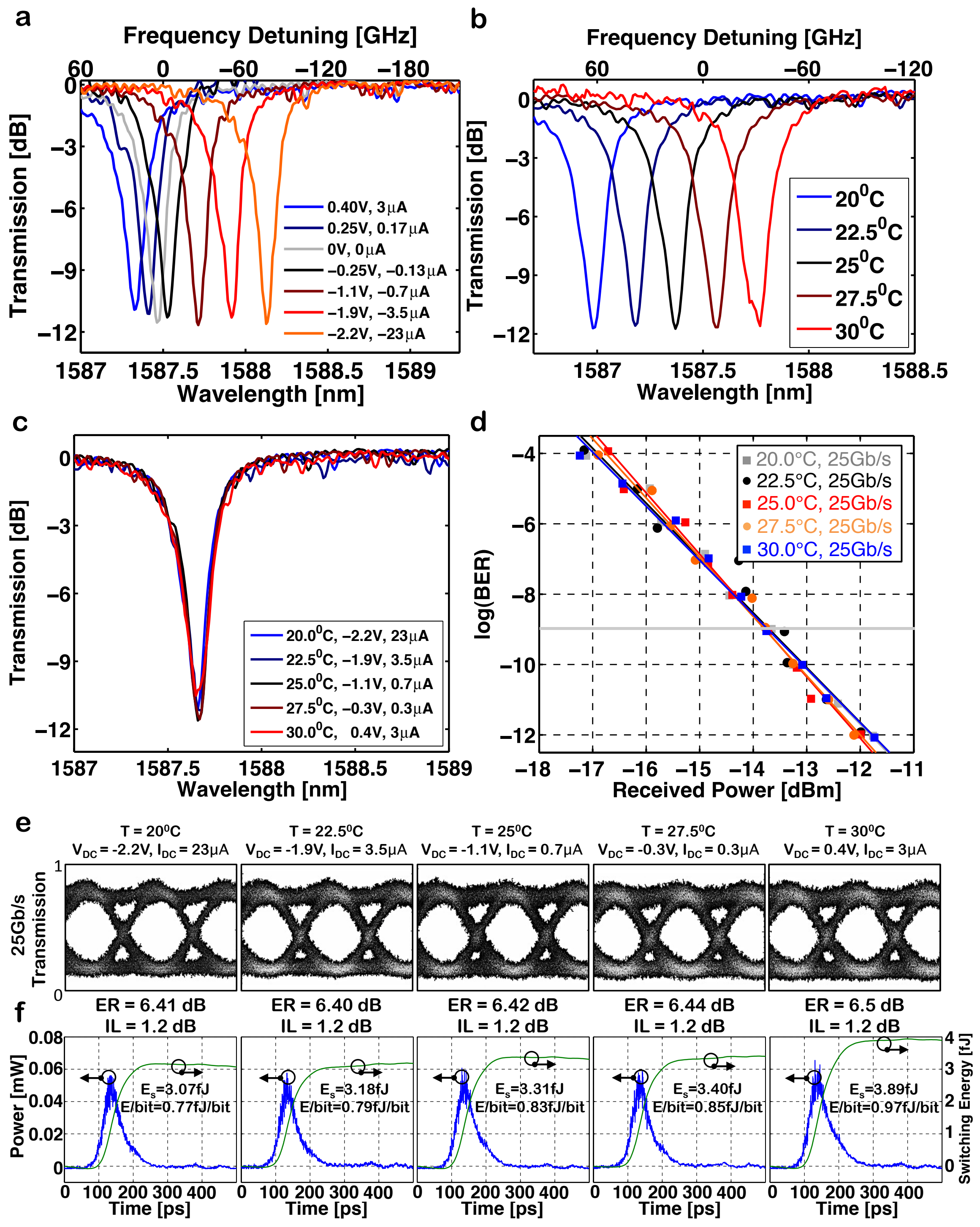

**Figure 4 | Athermal operation over 10°C temperature range. a**, The measured transmission spectra of the resonant modulator at ~26.5°C and applied DC bias voltages, ranging from 0.4V to -2.2V. **b,** The measured transmission spectra of the resonant modulator over a 10°C temperature range at 0V applied bias. The thermo-optic frequency response was extracted to be ~10GHz/°C. **c**, The measured transmission

spectra of the resonator over a 10°C temperature range with the modulator tuned electro-optically to compensate for thermal variations. **d**, The bit-error-rate curves measured over a 10°C temperature range for the silicon resonant modulator, which is driven with a terminated probe and $0.5V_{PP}$ NRZ-OOK PRBS with a pattern length of $2^{31}$-1 at a data rate of 25-Gb/s. The bit-error-rate curves clearly show error-free operation over the 10°C temperature range and the power penalty due to temperature variations is within ±0.05dB. **e**, Experimentally measured high-speed optical eye diagrams egressing from the silicon electro-optic modulator at a data rate of 25-Gb/s. The applied bias voltage ($V_{DC}$) and measured current ($I_{DC}$) are denoted above the eye diagrams at each temperature. The extinction ratio (ER) and insertion loss (IL) are denoted below the eye diagrams at each data rate and temperature. **f,** The TDR measurement of the electro-optic modulator power consumption and switching energy for a $0.5V_{PP}$ drive for the particular DC-bias voltages.

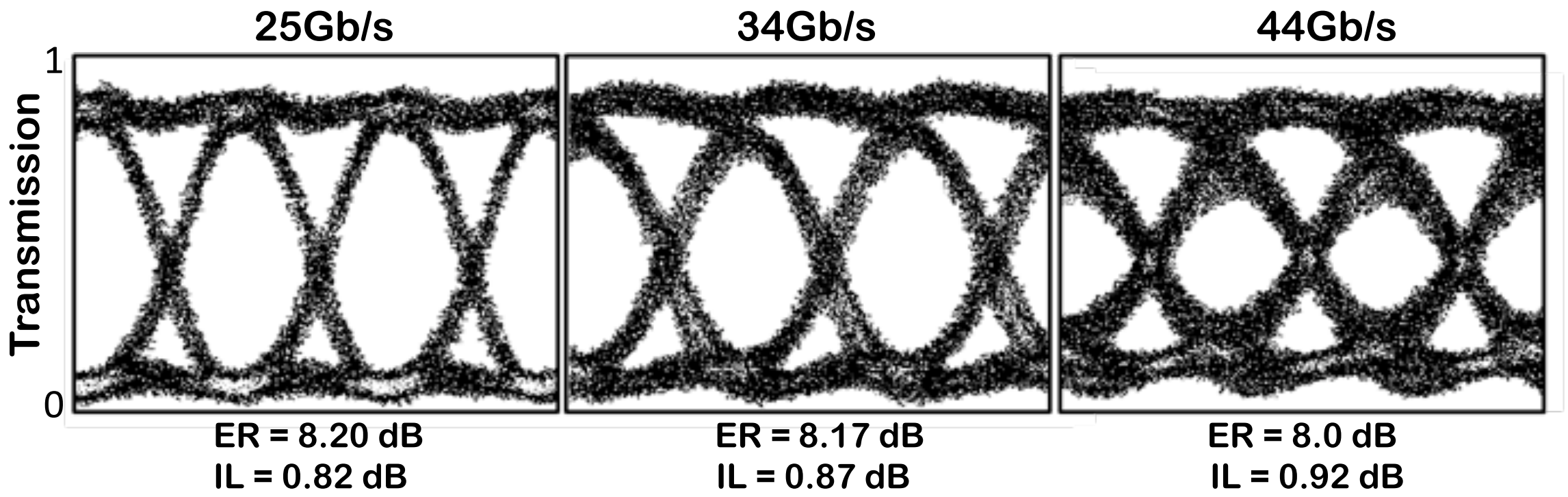

**Figure 5 | Bandwidth limited data transmission.** The experimentally measured high-speed optical eye diagrams egressing from the silicon microdisk modulator which is driven with a terminated probe and 2.2$V_{PP}$ NRZ-OOK PRBS with a pattern length of $2^{31}$-1 at data rates of 25-, 34- and 44-Gb/s. The extinction ratio (ER) and insertion loss (IL) are denoted below the eye diagrams at each data rate.